\documentstyle[preprint,aps]{revtex}

\begin{document}
\draft
\title{Theory of the Transition at 0.2 K in Ni-doped Bi$_2$Sr$_2$CaCu$_2$O$_8$}
\author{Robert Joynt}
\address{National Center for Theoretical Sciences \\
P.O. Box 2-131 \\
Hsinchu, Taiwan 300, Republic of China and\\
Department of Physics and Applied Superconductivity Center\\
University of Wisconsin-Madison \\
1150 University Avenue \\
Madison, WI 53706, USA \\
}
\date{\today}
\maketitle

\begin{abstract}
A theory is put forward that the electronic phase transition at 0.2 K in
Ni-doped Bi$_{2}$Sr$_{2}$CaCu$_{2}$O$_{8}$ is result of the formation of a
spin density wave in the system of Ni impurities. The driving force for the
transition is the exchange interaction between the impurity spins and the
spins of the conduction electrons. This creates a small gap at two of the
four nodes of the superconducting gap. The effect is to reduce the thermal
conductivity by a factor of two, as observed.
\end{abstract}


The gap structure of high-temperature superconductors should show up most
clearly at low temperatures, when the quasiparticles are located near the
nodes. Unfortunately, it is rather difficult to obtain reliable information
in this regime, because only relatively few bulk experiments are able to
probe properties of the particles and the effect of impurities on them.
Among these experiments, thermal conductivity $\kappa _{T}$ offers the most
accurate way of exploring the DC transport properties. Measurements of $%
\kappa _{T}$ in Zn-doped YBa$_{2}$Cu$_{3}$O$_{7}$ \cite{louis}have confirmed
the d-wave nature of the order parameter, while experiments in pure Bi$_{2}$%
Sr$_{2}$CaCu$_{2}$O$_{8}$ have shown anomalies in a field whose nature has
yet to be resolved \cite{ong}.

The effect of impurities on $\kappa _{T}$ is expected to be particularly
informative because of a well-developed theoretical machinery for their
calculation\cite{sun}. It was therefore particularly surprising that the
doped compound Bi$_{2}$Sr$_{2}$Ca(Cu$_{1-x}$Ni$_{x}$)$_{2}$O$_{8}$ has an
electronic phase transition beginning at a concentration of about $x=0.006$%
\cite{mov}. As the temperature is lowered through $T^{\ast }\cong 0.2K$, $%
\kappa _{T}$ drops by about a factor of two, indicating a sharp reduction in
the density of excited quasiparticles.

The explanation which has been offered for this sharp drop is a transition
to a state in which the superconducting order parameter breaks time-reversal
symmetry, the $(d_{x^{2}-y^{2}}+id_{xy})$ state\cite{sasha}. This
modification is caused by the spin-orbit interaction between the impurity
spins and the conduction electrons. In this paper I offer an alternative
explanation of the observations based on the exchange interaction between
the impurity spins and the conduction electrons which gives rise to a spin
density wave (SDW). I shall compare the two theories in the conclusion.

The Hamiltonian is

\begin{equation}
{\cal H} = {\cal H}_{d} - J \sum_i \vec{M}(\vec{R}_i) \cdot \vec{m}(\vec{R}%
_i),  \label{eq:exj}
\end{equation}
where ${\cal H}_{d}$ is the weak-coupling d-wave Hamiltonian and $\vec{M}(%
\vec{R}_i)$ is the spin of the Ni atom at the impurity site $i$ and $\vec{m}(%
\vec{R}_i)$ is the spin density of the conduction electrons at the same
point. This is the usual exchange interaction. It has been taken as local,
but this assumption is not crucial in what follows. Nor is the sign of $J$
important. The system is treated as two-dimensional.

I propose that there is spiral magnetic order in the impurity spins below $%
T^{\ast }$: 
\begin{equation}
\langle \vec{M}(\vec{R}_{i})\rangle =\vec{M}(T)\left[ \widehat{x}\cos (\vec{Q%
}\cdot \vec{R}_{i})+\widehat{y}\sin (\vec{Q}\cdot \vec{R}_{i})\right] .
\label{eq:m}
\end{equation}
Furthermore, take $\vec{Q}=(Q/\sqrt{2})(1,-1)$. The reason for this will
become clear below. Substitution of Eq.\ \ref{eq:m} into Eq.\ \ref{eq:exj}
shows that there is an effective magnetic field acting on the spins of
conduction electrons. It contains two terms. There is a coherent field with
wavevector $\vec{Q}$ and a random field arising from the fact that the set $%
\{\vec{R}_{i}\}$ is random. I concentrate on the first term here, since it
can produce the sudden drop in $\kappa _{T}$. The effects of the second term
are discussed below.

The energy density change in the conduction electron system is 
\begin{equation}
\Delta E(\vec{Q})=-\frac{J^{2}M^{2}n_{imp}}{8\mu _{B}^{2}}\sum_{\vec{k}}[S(%
\vec{k}+\vec{Q})-1]\chi _{m}(\vec{k}),
\end{equation}
where $S(\vec{k})$ is the static structure factor for the positions of the
impurities \cite{liquid}\ and $\chi _{m}(\vec{k}) $ is the susceptibility of
the conduction electrons. $n_{imp}$ is the two-dimensional density of
impurities. In order to determine $\vec{Q}$, this expression must be
minimized. The structure factor is that of a highly disordered solid. \ It
will have the peak at $\vec{k}=0$ required by a sum rule, and other
reciprocal lattice vectors (the first at $\left| \vec{k}\right| \sim \sqrt{%
n_{imp}}$) will be strongly suppressed by the Debye-Waller factor. The
susceptibility has an unusual structure in a d-wave superconductor: $\chi
_{m}(\vec{Q})=a\chi _{P}\xi _{0}|\vec{Q}|+\ldots +(T/\Delta _{m})\chi
_{P}(b+c\xi _{0}^{2}|\vec{{Q}}|^{2}+\ldots )$ plus terms higher-order in $T$%
. $\ a$,$b$, and $c$ are model-dependent constants, $\Delta _{m}$ is the
maximum gap at zero temperature, $\xi _{0}$ is the coherence length, and $%
\chi _{P}$ is the Pauli susceptibility.

At the very low temperatures of interest here, the non-analytic part is the
important one. Hence the product $\Delta E(\vec{Q})$ has a minimum at $|\vec{%
Q}_{0}|\sim \sqrt{n_{imp}}$, near the first non-zero peak in $S(\vec{k})$. \
Physically, the point is this. \ The dilute system of impurities can only
support waves whose wavelength is longer than the interimpurity spacing. \
The infinite-wavelength ferromagnetic state does not lower the energy of the
system because the bulk ferromagnetic susceptibility vanishes. \ Because $%
\chi _{m}(\vec{Q})\sim |\vec{Q}|$, the system chooses the shortest
wavelength that the dilute impurities will support. \ The ordering
wavevector $|\vec{Q}|$ is small, of order $0.04A^{-1} $. Observe also that
if the structure factor is entirely gas-like, then there is no transition.
There must be some short-range repulsion between the impurities.

A central point of this paper is that the non-analyticity peculiar to d-wave
systems is responsible for the transition. This phenomenon is presumably
related to the non-analyticity obtained in calculations of orbital
susceptibilties \cite{sauls}. \ In s-wave materials, the zero-temperature
susceptibility $\chi (\vec{q})$ starts with terms of order $|\vec{q}|^{2}$:
Eq. 1 shows that this will strongly suppress the energy of an SDW in a
disordered dilute impurity system. \ Of course in a dense sublattice of
magnetic atoms the situation is different\cite{phil} and SDW formation in
superconductors has been observed\cite{rev}.

Considering now the energy $\Delta E$ as an effective interaction in the
spirit of RKKY, we can calculate the critical temperature of the SDW in
mean-field theory for spin 1 on the Ni site: 
\begin{equation}
T^{\ast }=\frac{J^{2}[S(\vec{Q}_{0})-1]\chi _{m}(\vec{Q}_{0})}{12Ak_{B}\mu
_{B}^{2}}.
\end{equation}
Here $A$ is the area. \ Taking $[S(\vec{Q}_{0})-1]=0.01$, $T^{\ast }=0.2K$,
and a susceptibility of $\mu _{B}^{2}\times 1(eV)^{-1}$ per conduction
electron, we find a very reasonable value of $\ J\sim 50meV$ for the
exchange constant. \ However, because we lack precise knowledge of the
values of $J$ and $[S(\vec{Q}_{0})-1]$, we cannot use this formula to
compare with experiment. \ Fortunately, most of the unknowns in the theory
occur only in the product.

The SDW produces a change in the translation symmetry group. \ The first
Brillouin zone is now a thin slice in momentum space. \ Two of the nodes [
at $\pm (k_{F},k_{F})/\sqrt{2}$ ] are in the center of the short axis of the
zone and the other two [ at $\pm (k_{F},-k_{F})/\sqrt{2}$ ] are in
indeterminate position. \ Here $k_{F}$ is the length of the Fermi wavevector
along the diagonal. \ The states near the former point are strongly affected
by the spin ordering because pairs connected by the short reciprocal lattice
vector have small energy differences. \ The states near the other nodes that
connected by the short reciprocal lattice vector have large energy
differences: this vector is perpendicular to the Fermi surface. \ 

To see the effects on the thermal conductivity, we calculate the
quasiparticle energies in the presence of the coherent part of the effective
field in Eq.\ \ref{eq:exj}. The expression for these energies near a
magnetic zone boundary, obtained by diagonalizing the appropriate $4\times 4$
matrix, for this is: 
\begin{eqnarray}
{\cal E}_{\pm ,\pm }(k_{x},k_{y}) &=&\pm \frac{1}{2}\{E^{2}(\vec{k}+\vec{Q}%
_{0})+E^{2}(\vec{k})+2b_{eff}^{2}\pm \lbrack (E^{2}(\vec{k}+\vec{Q}%
_{0})+E^{2}(\vec{k}))^{2} \\
&&+16b_{eff}^{2}(\xi _{av}^{2}(\vec{k},\vec{Q}_{0})+\Delta _{av}^{2}(\vec{k},%
\vec{Q}_{0}))]^{1/2}\}^{1/2}.  \label{eq:eqp}
\end{eqnarray}
Here $b_{eff}=JxM$ and $E^{2}(\vec{k})=\xi ^{2}(\vec{k})+\Delta ^{2}(\vec{k}%
) $, where $\xi (\vec{k})$ are the normal-state quasiparticle energies
referred to the chemical potential and $\Delta (\vec{k})$ is the d-wave gap
function, while $\xi _{av}(\vec{k},\vec{Q}_{0})=[\xi (\vec{k}+\vec{Q}%
_{0})+\xi (\vec{k})]/2$ and $\Delta _{av}(\vec{k},\vec{Q}_{0})=[\Delta (\vec{%
k}+\vec{Q}_{0})+\Delta (\vec{k})]/2$. There is a level repulsion between the
particle-like and hole-like Bogoliubov branches. This leads to the
development of an energy minigap which is obtained from Eq.\ \ref{eq:eqp} by
setting $E(\vec{k})=\xi (\vec{k})=\Delta (\vec{k})=0$. The result for the
minigap energy for the nodes at $\pm (k_{F},k_{F})/\sqrt{2}$ is of order $%
b_{eff}^{2}/v_{s}Q_{0}$, where $v_{s}$ is the slope of the superconducting
energy gap at the node. $\ $It may be estimated as $v_{s}\sim \Delta
_{m}/k_{F}$, where $\Delta _{m}$ is the maximum value of the superconducting
gap. \ Then the minigap is roughly \ $10^{-2}meV$ when $M=1$ . \ The
corresponding gap at the other pair of nodes located at $\pm (k_{F},-k_{F})/%
\sqrt{2}$ is much smaller, of order $b_{eff}^{2}/v_{F}Q_{0}$, where $v_{F}$
is the Fermi velocity. The big minigap is of order $T^{\ast }$, while the
latter is smaller by a factor of about 30. As a result, the system develops
an appreciable gap in two of its nodes at the transition, while the other
two remain essentially ungapped at $T^{\ast }$.

The thermal conductivity may then be calculated in the presence of $M(T)$,
which is again obtained from spin one mean field theory. The result from the
standard Boltzmann equation approach is shown together with the data \cite
{mov} from a sample with $x=0.015$ in Fig.\ 1. The curve has been fit from $%
0.1K$ to $0.3K$ using as adjustable parameters the quantities $Q_{0}$, $%
T^{\ast }$, and $b_{eff}$ with the results $Q_{0}=24k_{B}T^{\ast }/v_{s}\sim
k_{F}/50$, $T^{\ast }=0.22K$, and $b_{eff}=8k_{B}T^{\ast }$. \ The fact that
the fit parameters are in good agreement with the {\it a priori }estimates
above is strong evidence for the correctness of the theory. \ To the extent
that the specific heaat can be measured, the entropy change in the same
temperature range changes by about $N_{imp} \ln 3$, which is consistent with
this picture.

The fit in Fig. 1 is poor at low temperatures owing to the neglect of the
random term in the effective field. It is known that this randomness leads
to a crossover from the clean result $\kappa _{T}\sim T^{2}$ at high
temperatures to $\kappa _{T}\sim T$ at low temperatures. This effect is
present in the data in Fig. 1, though in a temperature regime where the data
look noisy. \ The linear regime is not present in the theory as it stands.
Improvements in this direction are relatively straightforward and will be
presented in a longer paper. At higher temperatures the data follow a law $%
\kappa _{T}\sim T^{\alpha }$ with $\alpha \approx 1.6-1.75$, whereas the
current theory gives the usual result $\alpha =2$. \ There is no theory of
this intriguing observation at the present time. \ The corresponding
exponent in Zn-doped YBa$_{2}$Cu$_{3}$O$_{7}$ appears to greater than 2 \cite
{louis}.

Considered at the Ginzburg-Landau level, the present theory has the
structure: 
\begin{equation}
F=a_{M}(T)M^{2}-a_{Mm}Mm+a_{m}m^{2}+{\cal O}(M^{4},M^{2}m^{2},m^{4}).
\label{eq:f1}
\end{equation}
The order parameters $M$ and $m$ represent the amplitude of the SDW on the
impurities and on the conduction electrons respectively. The coefficient $%
a_{M}(T)$ is entirely of entropic origin, implying $a_{M}(T)\geq 0,$as the
direct interaction between the impurity spins is negligible $(\mu
_{B}^{2}n_{imp}^{3/2}/k_{B}\sim 10^{-4}K)$. \ The temperature dependence of
the other two coefficients $a_{Mm}$ and $a_{m}$ may be neglected, as they
are proportional to the exchange coupling and the inverse zero-temperaure
bulk susceptibility, respectively. \ The transition takes place when the
lowest eigenvalue of the quadratic form changes sign: this is the equation
for $T^{\ast }$ given above. \ This treatment makes it clear that the SDW
involves the ordering of both the impurity spins and the conduction
electrons. \ In this regard, the present picture resembles that put forward
by Balatsky \cite{sasha}, which postulates a time-reversal symmetry breaking
superconducting order parameter $(d_{x^{2}-y^{2}}+id_{xy})$. \ However, the
resemblance is superficial. \ In the $(d_{x^{2}-y^{2}}+id_{xy})$ theory, the
Ginzburg-Landau expansion has the 
\[
F=a_{M}(T)M_{z}^{2}+\alpha _{0}\left| \Delta _{0}\right| ^{2}+\alpha
_{1}\left| \Delta _{1}\right| ^{2}+ib_{M\Delta }M_{z}(\Delta _{0}\Delta
_{1}^{\ast }-\Delta _{0}\Delta _{1}^{\ast })+{\cal O}(M_{z}^{4},\Delta
_{0}^{4},\Delta _{1}^{4})
\]
Here $\Delta _{0}$ and $\Delta _{1\text{ }}$are the $d_{x^{2}-y^{2}}$ and $%
d_{xy}$ superconducting order parameters and $M_{z}$ is the amplitude of the
uniform magnetization on the impurity sites. \ The most important difference
from the free energy in Eq. \ref{eq:f1} lies in the fact that the coupling
term is trilinear. \ As a result, the phase transition can only arise from a
change of sign in $a_{M}(T)$. \ In view of the estimate of the direct
magnetic coupling given above, this is unlikely to occur at the temperatures
in question. \ The two theories differ also in their predictions for $\kappa
_{T}$. \ In the SDW theory, only one half of the nodes are substantially
gapped, leading to a natural explanation of the sudden drop by a factor of
two. \ In the $(d_{x^{2}-y^{2}}+id_{xy})$ theory all nodes are gapped. \ The
theory must rely on the formation of an impurity band of states in the gap t
explain the low temperature limit. \ This would imply that the
zero-temperature limit of $\kappa _{T}/T$ is extrinsic, depending on the
impurity concentration. \ The fact that \ $\kappa _{T}/T$ approaches a value
comparable to the universal limit makes this explanation unattractive, but
it is clearly consistent with the SDW theory. \ 

In closing it is appropriate to consider how to sharpen the comparison of
experiment and theory. \ The two-dimensional SDW transition of course needs
to be stabilized by interlayer coupling. \ It is not clear what the
threee-dimensional ordering will be, as the dipole or other coupling may
then compete with RKKY. \ In any case, the small impurity concentrations may
make the determination of the structure by neutron scattering difficult. \ A
further complication is the dependence of $T^{\ast }$ on the structure
factor. \ This can cause the transition temperature to depend on the
preparation method and annealing time. \ It could be responsible for the
fact that the transition temperatures in samples of Bi$_{2}$Sr$_{2}$Ca(Cu$%
_{1-x}$Ni$_{x}$)$_{2}$O$_{8}$ with nominal compositions $x=0.006$ and $%
x=0.015$ are about the same. \ It is likely that a different mechanism
suppresses the transition at $x=0.024$. \ At these higher concentrations $%
\chi _{m}(q=0,T=0)$ will become appreciable due to the random interaction
term. \ This will favor a transition into a ferromagnetic state, as there is
always a $\delta ^{2}(\vec{k})$ term present in $S(\vec{k})$. \ Repeating
the calculation of the quasiparticle energies as above shows that this does
not produce any gap in the quasiparticle spectrum and thus would not be seen
in $\kappa _{T}$. \ By the same token, application of a uniform magnetic
field produces a first-order transition to a ferromagnetic state and erases
the $\kappa _{T}$ anomaly, as is seen in the experiments. \ A simple way to
distinguish between the SDW and $(d_{x^{2}-y^{2}}+id_{xy})$ theories
experimentally is that the SDW responds isotropically to an applied field,
while the $(d_{x^{2}-y^{2}}+id_{xy})$ would predict the disappearance of the 
$\kappa _{T}$ anomaly only if the field is in the basal plane. \
Furthermore, the SDW would lead to transport anisotropy between the (1,1)
direction and the (1,-1) directions, if domain effects can be eliminated.
This could happen automatically if there is coupling of the SDW to the
superlattice distortion in the crystal structure.

I would like to thank T.K. Lee, B. Rosenstein, A. Knigavko, S.K. Yip, and
A.V. Balatsky for helpful discussions. This work was supported by the NSF
under the Materials Theory Program (DMR-9704972) and under the Materials
Research Science and Engineering Center Program (DMR-96-32527). \ \

\begin{figure}[tbp]
\caption{Comparison of theory(solid line) and experimental data (points)for
the thermal conductivity divided by temperature versus temperature. The data
are taken from Ref.\ 4.}
\label{fig:expdat}
\end{figure}

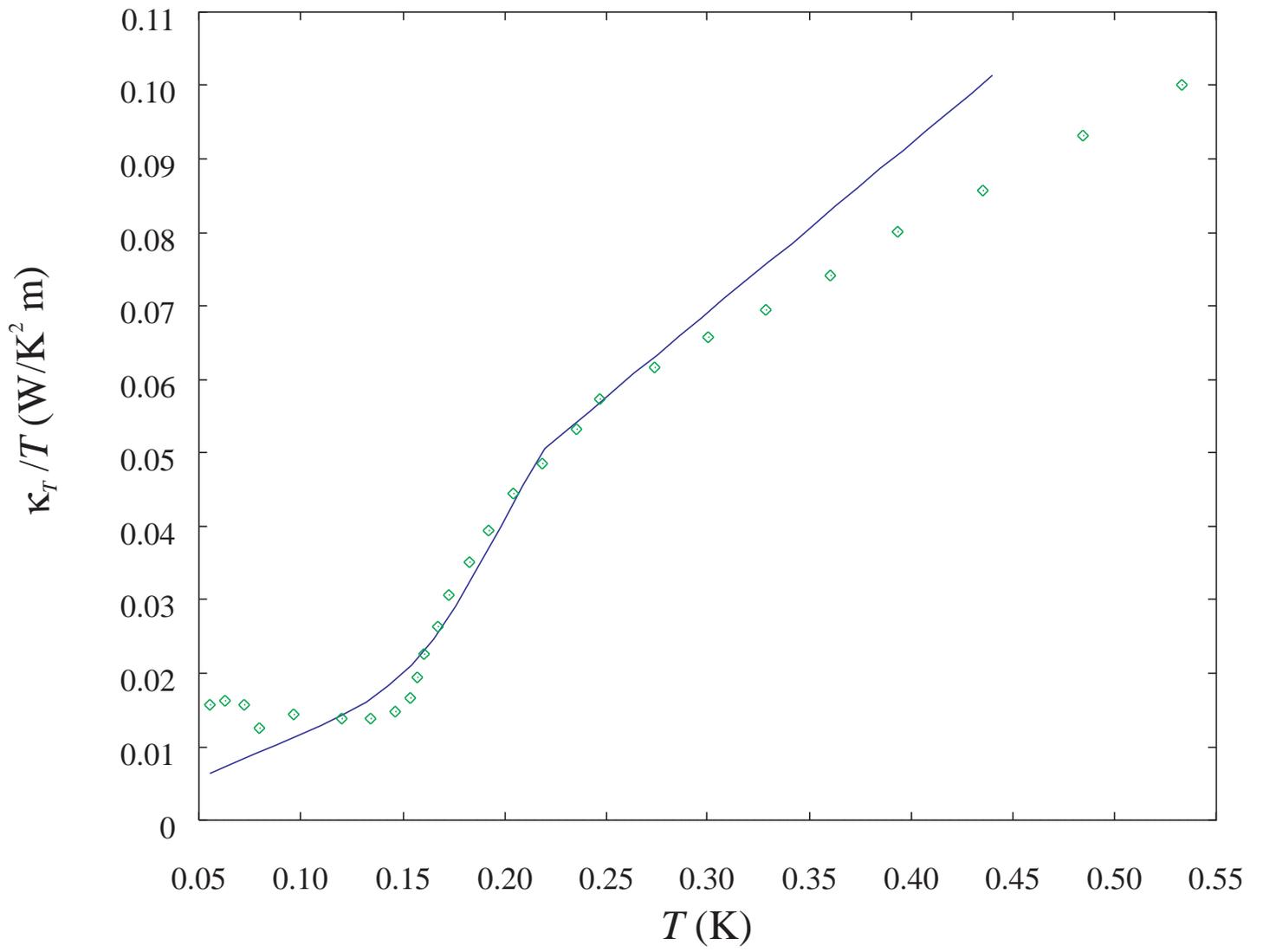
\end{document}